\documentclass[runningheads]{llncs}
\usepackage{graphicx}
\usepackage{booktabs}
\usepackage{diagbox} 
\usepackage{multirow}
\usepackage[dvipsnames]{xcolor}
\usepackage[misc]{ifsym}
\begin{document}

\title{Three-Dimensional Embedded Attentive RNN (3D-EAR) Segmentor for Left Ventricle Delineation from Myocardial Velocity Mapping}

\titlerunning{3D-EAR for MVM-CMR}

\author{Mengmeng Kuang\inst{1} \Letter, Yinzhe Wu\inst{2,3}, Diego Alonso-Álvarez\inst{4}, \\ David Firmin\inst{2,5}, Jennifer Keegan\inst{2,5}, Peter Gatehouse\inst{2,5}, \\ and  Guang Yang\inst{2,5} \orcidID{0000-0001-7344-7733} \Letter}

\authorrunning{M. Kuang et al.}

\institute{$^1$ Department of Computer Science, The University of Hong Kong\\
$^2$ National Heart and Lung Institute, Imperial College London\\
$^3$ Department of Bioengineering, Imperial College London\\
$^4$ Research Computing Service, Information and Communication Technologies, \\ Imperial College London\\
$^5$ Cardiovascular Research Centre, Royal Brompton Hospital\\
\email{mmkuang@cs.hku.hk} and \email{g.yang@imperial.ac.uk}
}

\maketitle              

\begin{abstract}
Myocardial Velocity Mapping Cardiac MR (MVM-CMR) can be used to measure global and regional myocardial velocities with proved reproducibility. Accurate left ventricle delineation is a prerequisite for robust and reproducible myocardial velocity estimation. Conventional manual segmentation on this dataset can be time-consuming and subjective, and an effective fully automated delineation method is highly in demand. By leveraging recently proposed deep learning-based semantic segmentation approaches, in this study, we propose a novel fully automated framework incorporating a 3D-UNet backbone architecture with \textbf{\underline{E}}mbedded multichannel \textbf{\underline{A}}ttention mechanism and LSTM based \textbf{\underline{R}}ecurrent neural networks (RNN) for the MVM-CMR datasets (dubbed 3D-EAR segmentor). The proposed method also utilises the amalgamation of magnitude and phase images as input to realise an information fusion of this multichannel dataset and exploring the correlations of temporal frames via the embedded RNN. By comparing the baseline model of 3D-UNet and ablation studies with and without embedded attentive LSTM modules and various loss functions, we can demonstrate that the proposed model has outperformed the state-of-the-art baseline models with significant improvement.

\end{abstract}

\section{Introduction}

Healthy functioning of the left ventricle requires complex motion and all parts of the myocardium must perform synergistically in order to ensure the heart to pump efficiently. In certain pathologies, early regional myocardial instability may be compensated for by altered movement in other areas in order to maintain the ventricular function. Global myocardial velocities are widely used in clinical practice; however, the global measurement might only be detectable when the condition has advanced to a point where compensation is no longer possible. By additional measurement of local myocardial dynamics, myocardial stability can be more specifically quantified and potential cardiovascular disease can therefore be detected earlier \cite{simpson2014spiral}.

Among different cardiac MR (CMR) techniques, there are a few methods that can be used to calculate global and regional myocardial dynamics \cite{simpson2013mr}. Myocardial Velocity Mapping CMR (MVM-CMR) \cite{simpson2013efficient} can potentially provide both high spatial and temporal resolution, and has clear advantages compared to the blood velocity scans that are commonly used in clinical.

Accurate segmentation of the left ventricle (LV) is the first step and a prerequisite for robust and reproducible global and regional myocardial velocity estimation. Conventional manual segmentation on the MVM-CMR dataset can be extremely time-consuming considering both the high spatial and temporal resolution of the dataset. Such manual segmentation is also limited by clinician's experience and potential human operator fatigue may also affect the delineation accuracy. Therefore, an efficient and robust automated LV segmentation method for the multichannel (i.e., magnitude and three velocity channels) and multi-frame (i.e., temporal frames of the cardiac "movie" acquired) MVM-CMR data is necessary for the clinical deployment of the global and regional velocities estimation.


Development in deep learning represents a major leap for digital healthcare. Several research studies have demonstrated promising results for the anatomical and pathological segmentation of the heart from CMR images, e.g., whole heart segmentation \cite{zhuang2019evaluation}, LV segmentation \cite{bai2018automated}, left atrial segmentation \cite{yang2020simultaneous} and atrial scar delineation \cite{li2020atrial}. A more detailed review of the segmentation of cardiac images can be found elsewhere \cite{chen2020deep}.

Despite successful applications of deep learning based techniques for CMR data segmentation, a plain deployment of existing methods for the MVM-CMR data can be challenging, but more informative. This is due to (1) multi-frame temporal MVM-CMR data may require a more complicated network design to ensure both accurate slice-wise delineation and continuity in the temporal dimension and (2) MVM-CMR data have multiple channels, e.g., magnitude and phase channels, that can provide richer information of the LV anatomy, but how to explore such informative multichannel data is still an open question.

Inspired by recent progress on semantic segmentation, e.g., UNet \cite{ronneberger2015u}, we propose a novel \textbf{\underline{E}}mbedded multichannel \textbf{\underline{A}}ttentive \textbf{\underline{R}}ecurrent neural networks (RNN), abbreviated 3D-EAR segmentor, for LV delineation from MVM-CMR datasets. The proposed 3D-EAR segmentor consists of three major components: (1) a 3D-UNet based backbone network, (2) embedded attention modules to enhance the network skip connections for more accurate localisation of the LV anatomy, and (3) long short-term memory (LSTM) based RNN modules to learn the temporal information of the multi-frame context at the bottom of the U-shaped network. In doing so, the proposed method can leverage the amalgamation of magnitude and phase images as more informative input to realise an effective information fusion of this multichannel dataset. Besides, the temporal dimension continuity can be ensured by exploring the correlations of temporal frames via the embedded LSTM. In addition, by varying different loss functions (i.e., Cross-Entropy loss, Dice loss \cite{milletari2016v}, and Dice-IoU \footnote{IoU stands for Intersection-Over-Union, which is also known as the Jaccard index.} loss \cite{rahman2016optimizing}), we perform ablation studies to find the optimal architecture of the proposed network.

By validation on MVM-CMR data collected from healthy controls, our proposed 3D-EAR segmentor achieves superior LV segmentation performance compared to state-of-the-art baseline models at the patient-level. 


\section{Method}

\subsection{Data Acquisition, Preprocessing and Augmentation} \label{sub:data}

\begin{figure}[htbp]
\centering
\includegraphics[width=12cm]{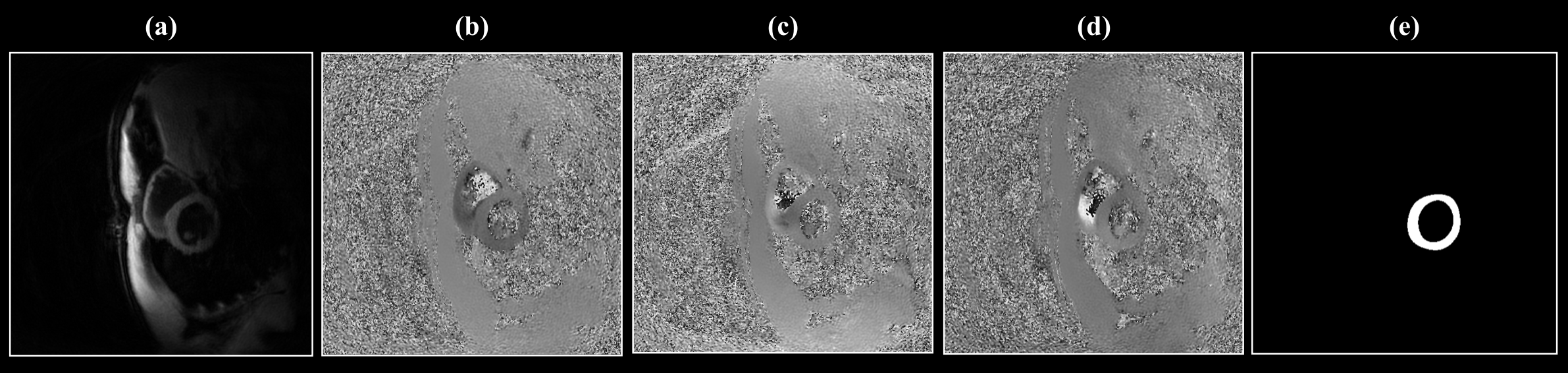}
\caption{A sample frame of our multi-channel MVM-CMR dataset with the manual segmentation. From left to right: magnitude channel, three phase channels and the manual LV segmentation.}
\label{fig:channels}
\end{figure}

The training and testing datasets contain 26 MVM-CMR datasets with the data size of 50 $\times$ 512 $\times$ 512 $\times$ 4 which were acquired from 18 healthy subjects (8 of them were acquired twice, giving 26 datasets) at Royal Brompton Hospital. Each of the datasets consists of 3-5 cine slices, giving 121 cine slices in total. There are 50 temporal frames per cardiac cycle and 4 channels reconstructed by a non-Cartesian SENSE reconstruction channel (one magnitude encoding channel and three velocity encoding channels of orthogonal directions), constituting the multi-frame multichannel MVM-CMR data. The MVM-CMR slices have spatial resolutions of 0.85mm $\times$ 0.85mm that were reconstructed from the acquired 1.7mm $\times$ 1.7mm. The MVM-CMR were acquired in short-axis slices from base to apex of the LV. An experienced cardiac MRI physicist performed manual delineation of the LV myocardium to create the ground truth for this study. In addition, we augmented the data by random rotation (angle =[90$^\circ$, 180$^\circ$, 270$^\circ$]) before model training. An example of our multichannel MVM-CMR dataset with the manual segmentation can be found in Figure \ref{fig:channels}.

\subsection{Network Architectures} \label{subsec:structures}

\subsubsection{Attention Enhanced Skip Connections}

We propose embedded attention modules to realise the enhanced skip connections and synthesise features from the transferred image $F$ of the convolutional layers in the original 3D-UNet structure. The objective is that the relevant features in a single slice can be enhanced during the attention process, while the less relevant ones can be less focused on. To achieve this, an attention map is computed in every frame of the input MVM-CMR images to represent the confidence of the transferred features for each position as expressed in Equation (\ref{equ:atten}) as follows
\begin{equation} \label{equ:atten}
    \mathrm{Layer}_{i}^{\mathrm{Attention}} = \mathrm{Softmax}(F_{i} \times \mathrm{Conv}(F_{i}))) \cdot \mathrm{Conv}(F_{i})', 
\end{equation}
where $i$ donates the $i$-th frame in the 3D feature map. $\mathrm{Conv}$ and $\mathrm{Softmax}$ represent the convolutional layer and the Softmax activation operation, respectively (Figure \ref{fig:lstm-atten} (a)).

\subsubsection{LSTM Based Temporal Feature Extractor}

We also develop LSTM layers to capture the cross-frame features at the bottom of the U-shaped structure in the 3D-UNet network. We assume that the LSTM can learn the temporal correlations from the multi-frame MVM-CMR data. For our 3D (2D+t) MVM-CMR data, we need to convert them into sequences and then transfer back into 3D (i.e., 2D+t) images before and after the LSTM layer. The whole operation can be denoted as Equation (\ref{equ:lstm}) that is
\begin{equation} \label{equ:lstm}
    \mathrm{Layer}_{\mathrm{out}} = \mathrm{Reshape}(\mathrm{LSTM}(\mathrm{Reshape}(\mathrm{Layer}_{\mathrm{in}}))).
\end{equation}

\begin{figure}[!htb]
    \centering
    \includegraphics[height=4.5cm]{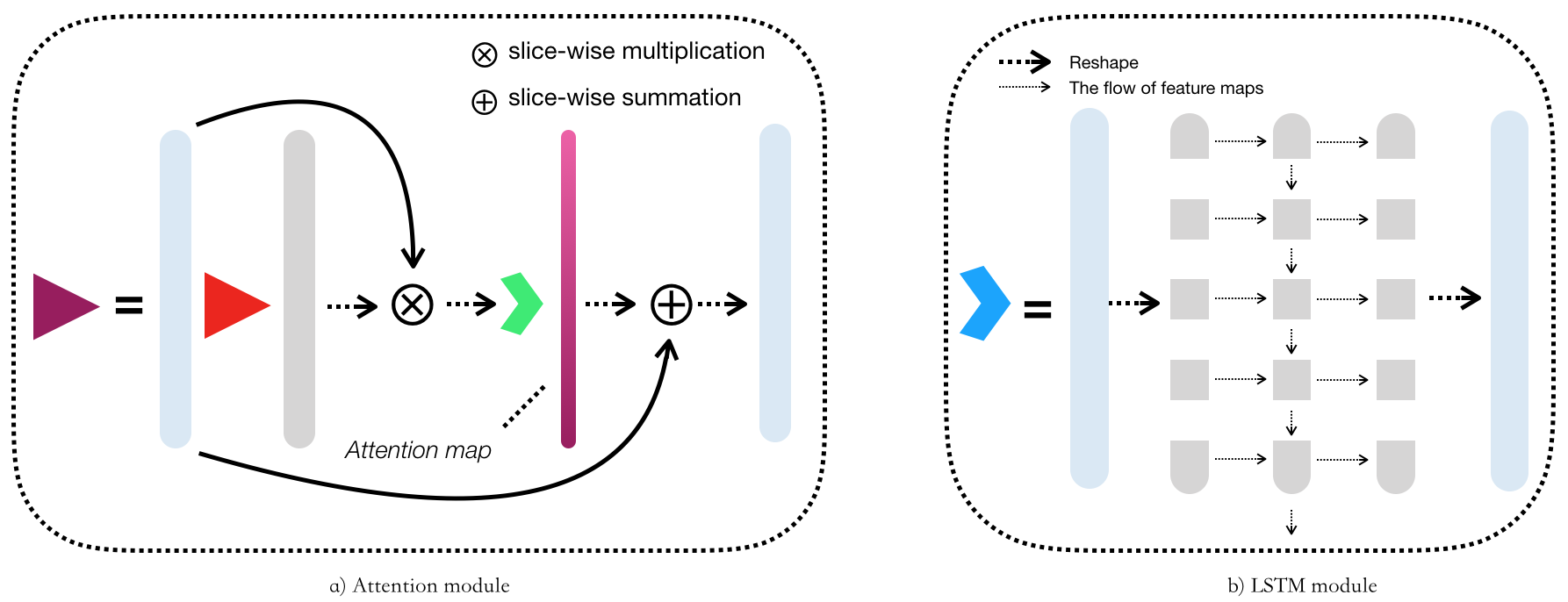}
    \caption{The structure of the embedded in-slice attention block (a) and cross-frame LSTM module (b).}
    \label{fig:lstm-atten}
\end{figure}

\begin{figure}[!htb]
    \centering
    \includegraphics[width=12cm]{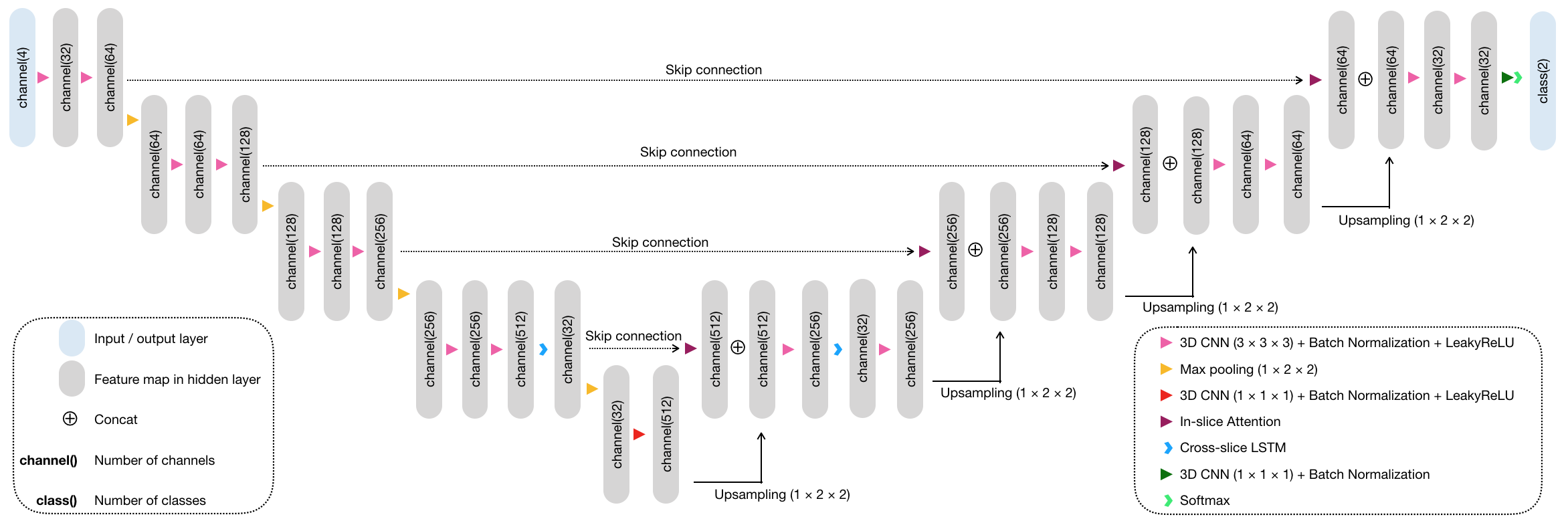}
    \caption{The overall architecture of our proposed 3D-EAR segmentor.}
    \label{fig:frame}
\end{figure}

Figure \ref{fig:lstm-atten} (b) shows the LSTM workflow and Figure \ref{fig:frame} represents the overall structure of the proposed model (i.e., 3D-EAR segmentor). In this figure, we use rounded rectangles to denote the flow of feature maps, triangles of different colours to represent different neural network blocks and arrows to indicate the skip connections and up-sampling.

\subsection{Loss Functions} \label{subsec:loss}

For our 3D-EAR segmentor, we implement various loss functions, e.g., (1) Cross-Entropy loss, (2) Dice loss (with Laplace smoothing) and (3) Dice-IoU loss (with Laplace smoothing) to seek an optimal solution. The standard Cross-Entropy loss can be represented by Equation (\ref{equ:celoss}), that is
\begin{equation}
\label{equ:celoss}
    \mathrm{Loss}_{\mathrm{Cross-Entropy}} = - \frac{1}{n} \sum_{i}\sum_{c=1}^{n} y_{i c} \log \left(p_{i c}\right),
\end{equation}
where $y$ and $p$ represent the true and predicted labels in the $n^{\mathrm{th}}$ class, respectively.

The Dice loss (with Laplace smoothing factor $f_{\mathrm{smooth}}$) can be denoted as Equation (\ref{equ:lossdice}), that is
\begin{equation} \label{equ:lossdice}
    \mathrm{Loss}_{\mathrm{Dice}} = 1 - \frac{2 \times |\mathrm{GT} \cap \mathrm{Pred}| + f_{\mathrm{smooth}}}{|\mathrm{GT}| + |\mathrm{Pred}| + f_{\mathrm{smooth}}},
\end{equation}
where $\mathrm{GT}$ stands for the ground truth of the segmentation, $\mathrm{Pred}$ donates the prediction of the model and $|\bullet|$ represents the area of $\bullet$.

The Dice-IoU loss, which is a combination of the Dice loss with the IoU calculation, can be represented as Equation (\ref{equ:lossall})
\begin{equation} \label{equ:lossall}
    \mathrm{Loss}_{\mathrm{Dice-IoU}} = \frac{1}{n} \times \sum_{i = 1}^{n} {\mathrm{Loss}_{\mathrm{Dice}} \times \mathrm{IoU}}.
\end{equation}
Equation (\ref{equ:lossiou}) illustrates the IoU calculation, which is also smoothed by the Laplace factor $f_{\mathrm{smooth}}$, that is
\begin{equation} \label{equ:lossiou}
    \mathrm{IoU} = 1 -  \frac{|\mathrm{GT} \cap \mathrm{Pred}| + f_{\mathrm{smooth}}}{|\mathrm{GT} \cup \mathrm{Pred}| + f_{\mathrm{smooth}}}.
\end{equation}

\subsection{Implementation Details} \label{subsec:impl}

The input of our model (and compared models) was a 4-channel 50-frame 512 $\times$ 512 MVM-CMR dataset, and the output prediction was with the same size as the input but had 2 different labels (i.e., LV and non-LV). We divided the MVM-CMR cine slices into two sets for experiments, one consisting of 80\% of the subjects for model training and the other one consisting of the remaining 20\% as independent testing. During the training process, we also performed the 5-fold cross-validation. The training was carried out on two standard NVIDIA GEFORCE RTX 2080 Ti GPUs. Our implementation was based on Keras and TensorFlow backend. The implementation and pre-trained models will be open source (on Github) for a reproducible study.

\section{Results}

\subsection{Experiments and Evaluation Metrics} \label{subsec:measure}

We performed the following comparison and ablation studies, including UNet3D (the baseline model 3D-UNet), UNet3D-Attention (3D-UNet with attention) and our proposed 3D-EAR segmentator with various loss functions. We evaluated model performance using (1) Dice scores, (2) Sensitivities and (3) Positive Predictive Values (PPV).

\subsection{Quantitative Results}

Quantitative results of our comparison study can be found in Table \ref{tab:dice}.
\begin{table}[!htb]
    \centering
        \caption{Dice scores of our comparison studies and ablation studies using various loss functions.}
    \begin{tabular}{r|p{2.5cm}|p{2.5cm}|p{2.5cm}}
    \toprule
        \diagbox{Structures}{Losses} & Cross-Entropy & Dice  & Dice-IoU \\
        \hline
        UNet3D & 0.84$\pm{0.02}$ & 0.85$\pm{0.03}$ & 0.88$\pm{0.02}$ \\
        \hline
        UNet3D-Attention & 0.87$\pm{0.03}$ & 0.87$\pm{0.02}$ & 0.89$\pm{0.02}$ \\
        \hline
        3D-EAR & 0.88$\pm{0.03}$ & 0.89$\pm{0.02}$ & 0.91$\pm{0.03}$\\
        \bottomrule
    \end{tabular}
    \label{tab:dice}
\end{table}


Table \ref{tab:dice} and Table \ref{tab:sen-ppv} show outstanding segmentation performance of using our proposed 3D-EAR model. Compared to the baseline model, our proposed 3D-EAR has achieved significantly higher Dice scores, sensitivities and PPV. We can also find that with the LSTM, our 3D-EAR has further improvement on the model with only the attention module.  



\begin{table}[!htb]
    \centering
        \caption{Sensitivities and PPV of our comparison studies and ablation studies using various loss functions.}
    \begin{tabular}{r|p{1.5cm}|p{1.5cm}|p{1.5cm}|p{1.5cm}|p{1.5cm}|p{1.5cm}}
    \toprule
    \multirow{2}*{\diagbox{Structures}{Losses}} & \multicolumn{2}{|c}{Cross-Entropy} & \multicolumn{2}{|c}{Dice} & \multicolumn{2}{|c}{Dice-IoU}\\
    \cline{2-7}
    &Sensitivity & PPV & Sensitivity & PPV & Sensitivity & PPV \\
        \hline
        UNet3D & 0.75$\pm{0.03}$ & 0.95$\pm{0.02}$ & 0.81$\pm{0.04}$ & 0.90$\pm{0.05}$ & 0.86$\pm{0.03}$ & 0.91$\pm{0.01}$ \\
        \hline
        UNet3D-Attention & 0.84$\pm{0.02}$ & 0.90$\pm{0.02}$ & 0.84$\pm{0.01}$ & 0.91$\pm{0.02}$ & 0.85$\pm{0.02}$ & 0.93$\pm{0.02}$\\
        \hline
        3D-EAR & 0.80$\pm{0.01}$ & 0.98$\pm{0.01}$ & 0.86$\pm{0.01}$ & 0.93$\pm{0.02}$ & 0.87$\pm{0.01}$ & 0.96$\pm{0.02}$ \\
        \bottomrule
    \end{tabular}
    \label{tab:sen-ppv}
\end{table}



\begin{figure}[!htb]
\centering
\includegraphics[width=12cm]{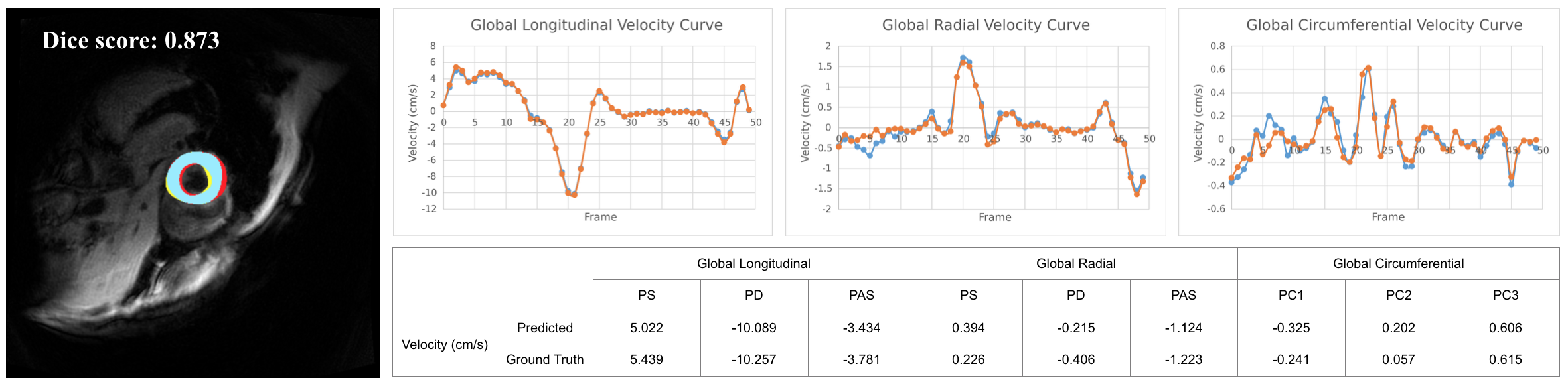}
\caption{A typical example of the segmentation results randomly selected from our MVM-CMR datasets with the global longitudinal, radial and circumferential velocity curves and peak velocities per slice of the example frames (More examples of full cardiac cycle segmentations will be provided in the Supplementary Material). For the segmentation: Blue---true positive; Yellow---false positive; Red--- false negative; Blue and Yellow regions---automated segmentation results. For the global velocity curves: Blue/Orange curves---derived from automated and manual segmentations.}
\label{fig:results}
\end{figure}

An automated segmentation example result obtained by our 3D-EAR model is shown in Figure \ref{fig:results}.
Followed by a morphological post-processing stage, we were able to generate their LV myocardium global velocity curves from the predicted results and compare it with the ones derived from the ground truth. We are able to observe close alignments of curves and little differences in the peak velocities generated from these curves.

\section{Discussion and Conclusion}

In this study, we have developed and validated a novel 3D-EAR segmentor for the delineation of LV from MVM-CMR data. The proposed model incorporated embedded attention enhanced skip connections to filter our irrelevant features from images and LSTM based RNN for accounting correlations among temporal frames of the MVM-CMR data. The experimental results have shown promising quantification and visualisation that can facilitate accurate and reliable estimation of global and local myocardial velocities. More detailed method descriptions, comparison results with and without multichannel input data, ablation studies of network parameters, and velocity comparisons in all the slices will be presented. 



\section{Acknowledgement}

This study was supported in part by BHF (TG/18/5/34111, PG/16/78/32402), in part by Heart Research UK RG2584, in part by the ERC IMI [101005122], and in part by ERC H2020 [952172].


\bibliographystyle{unsrt}
\bibliography{bibliography.bib}

\end{document}